\documentclass[aps,prl,twocolumn,noshowpacs,preprintnumbers,amsmath,amssymb]{revtex4-1}
\usepackage{graphicx}
\usepackage{bm}
\usepackage{upgreek}

\begin{document}

\title{Influence of the Fermi surface geometry on a Josephson effect between an iron-pnictide and conventional superconductors}

\author{A. A. Kalenyuk$^{1,2}$}
\author{E.~A.~Borodianskyi$^1$}
\author{A. A. Kordyuk$^{2,3}$}
\author{V. M. Krasnov$^{1,4}$}
\email{Vladimir.Krasnov@fysik.su.se}

\affiliation{$^1$Department of Physics, Stockholm University,
AlbaNova University Center, SE-10691 Stockholm, Sweden; }

\affiliation{$^2$ Institute of Metal Physics of National Academy
of Sciences of Ukraine, 03142 Kyiv, Ukraine;}

\affiliation{$^3$ Kyiv Academic University, 03142 Kyiv, Ukraine;}

\affiliation{$^4$ Moscow Institute of Physics and Technology,
State University, 141700 Dolgoprudny Russia.}

\begin{abstract}
We study Josephson junctions between a multi-band iron-pnictide
Ba$_{1-x}$Na$_x$Fe$_2$As$_2$
and conventional $s$-wave superconductors Nb and Cu/Nb bilayer. 
We observe that junctions with a Cu interlayer exhibit much larger
$I_c R_n$, despite a weaker proximity-induced superconductivity.
This counterintuitive result is attributed to the difference in
Fermi surface geometries of Nb and Cu,
which leads to a selective one-band tunneling from Cu and a
non-selective multi-band tunnelng from Nb. The latter leads to a
mutual cancellation of supercurrents due to the sign-reversal
$s_{\pm}$ symmetry of the order parameter in the pnictide. Our
results indicate that Fermi surface geometries play a crucial role
for pnictide-based junctions. This provides a new tool for phase
sensitive studies and paves a way to a conscious engineering of
such junctions.

\end{abstract}


\maketitle

Electronic structure of superconductors is usually quite
complicated, even for low-$T_c$ materials, such as the transition
metal Nb. Nevertheless, a simple description of Josephson effects,
which does not take into account complex Fermi surface (FS)
geometry, works remarkably well for conventional superconductors
\cite{Barone,Likharev}. This happens because probabilities of
electron and Cooper-pair tunneling are similar
\cite{McDonald_2001}. Together with a momentum-independent
$s$-wave energy gap, $\Delta$, it leads to the inverse
relationship between the normal resistance, $R_n$, and the
critical current, $I_c$. Thus, the $I_c R_n$ product becomes a
universal function of $\Delta$, independent of FS geometry
\cite{AB_1963}.

This universality, however, breaks for unconventional multi-band
superconductors. A particularly drastic deviation should occurs in
the case of sign-reversal order parameter
\cite{Tanaka_1997,Ota_2009,Linder_2009,Burmistrova_2015}. This
occurs in cuprate and iron-based superconductors, which are
believed to have $d$-wave \cite{vHarlingen_1993,Tsuei2000} and
$s_{\pm}$ \cite{Ng_2009,HirschfeldRPP2011,Wang2011,ChubukovAR2012}
symmetries, respectively. In this case, $I_c$ depends on gap
values in each band, and the $I_c R_n$ is band-structure-sensitive
and not universal
\cite{Tanaka_1997,Ota_2009,Linder_2009,Burmistrova_2015,Suppl}.


In this work we fabricate and study high-quality Josephson
junctions (JJ's) between single crystals of an iron-pnictide
Ba$_{1-x}$Na$_x$Fe$_2$As$_2$ (BNFA) and conventional low-$T_c$
superconductors made of either Nb film or Cu/Nb bilayer. Both
types of JJ's exhibit clean and clear Josephson phenomena.
However, JJ's with a Cu interlayer exhibit almost two order of magnitude 
larger $I_c R_n$, despite a weaker proximity-induced
superconductivity in Cu. This counterintuitive result is
attributed to the difference in FS geometries of Nb [multiple FS's
at various parts of the Brillouin zone (BZ)] and Cu [a single
quasi-spherical FS]. Therefore, tunneling from Nb takes place into
all bands of BNFA. Due to the sign-reversal $s_{\pm}$ order
parameter in BNFA, this leads to a mutual cancellation of
supercurrents and a very small $I_c R_n\sim 3 \mu$V. To the
contrary, tunneling from Cu occurs predominantly into one sub-band
avoiding such cancellation and leading to a significantly larger
$I_c R_n\simeq 200~ \mu$V. Our results indicate that
FS geometries play a crucial role for JJ's with multi-band,
sign-reversal superconductors. This provides a new tool for
fundamental studies of unconventional superconductivity and opens
a possibility for optimization and adjustment of junction
characteristics.

\begin{figure*}[t]
    \centering
    \includegraphics[width=0.99\textwidth]{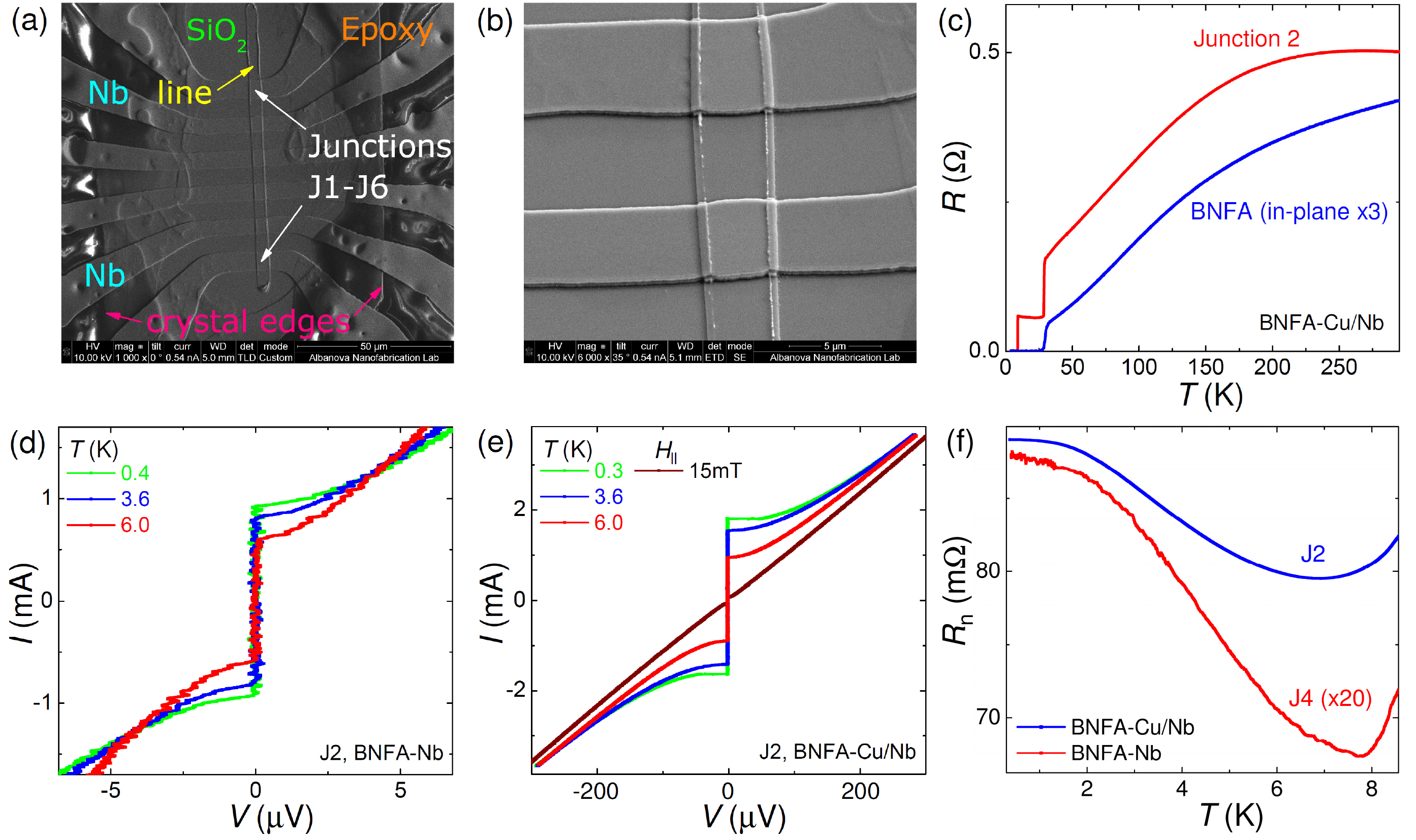}
    \caption{(Color online). (a) and (b) SEM images of the BNFA-Cu/Nb
    sample. The sample contains six junctions J1-6 (a), formed at
    a cross-like overlap between a window in the insulation SiO$_2$ layer and
    top electrodes (b). (c) Resistive transitions of a junction (J2,
    red) and BNFA crystal (blue) for the BNFA-Cu/Nb sample. (d)
    and (e) $I$-$V$ curves at different temperatures and $H=0$ for junctions
    at (d) BNFA-Nb and (e) BNFA-Cu/Nb samples. The wine curve in
    (d) represents $I$-$V$ at $T\simeq 0.3$ K and $H_{\parallel}=15$
    mT. It demonstrates a complete suppression of $I_c$ by a modest in-plane magnetic field.
    (f) Temperature dependencies of normal resistances for
    BNFA-Nb (red) and BNFA-Cu/Nb (blue) junctions.
     }
     \label{fig:fig1}
\end{figure*}


Figure \ref{fig:fig1} (a) represents a scanning electron
microscope (SEM) image of the BNFA-Cu/Nb sample. Our samples
contain six junctions made on a freshly cleaved BNFA single
crystal. Fig. \ref{fig:fig1} (b) shows a closeup on the junction.
Here the vertical strip represents the window in SiO$_2$ isolation
layer and the horizontal strip - the top contact electrode.
Micrometer-size JJ's are formed at the overlap between the two
strips. As the top electrode we use either pure Nb film ($\sim
200$ nm thick) or Cu(15 nm)/Nb(180 nm) bilayer deposited by
magnetron sputtering in a single cycle without breaking vacuum.
Details of sample fabrication, experimental setup and a list of JJ
parameters can be found in the Supplementary \cite{Suppl}.

Multiterminal configuration of our samples allows simultaneous
measurements of junction and crystal characteristics
\cite{Kalenyuk_2017,Kalenyuk_2018}. The blue line in Fig.
\ref{fig:fig1} (c) shows the in-plane resistive transition of
BNFA. At $T\sim 150$ K there is a kink in $R(T)$, corresponding to
a structural transition and spin-density-wave (SDW) ordering
\cite{HirschfeldRPP2011,ChubukovAR2012}. The superconducting
transition occurs at $T_c(BNFA)\simeq 30$ K. Observation of the
SDW state and the slightly sub-optimal $T_c$ indicate that the
BNFA crystal is moderately underdoped. The red line in Fig.
\ref{fig:fig1} (c) shows a simultaneously measured resistive
transition of a junction. It has two steps, first at $T_c(BNFA)$
and the second at $T_c(Nb)\sim 9$ K.

Figs. \ref{fig:fig1} (d) and (e) show current-voltage ($I$-$V$)
characteristics at different $T$ for (d) BNFA-Nb and (e)
BNFA-Cu/Nb JJ's. In both cases the $I$-$V$'s have the shape
typical for resistively shunted JJ's \cite{Barone,Likharev} with a
well defined $I_c$ and $R_n$. Green, blue and red curves are
measured at zero field. The wine-color line in Fig. \ref{fig:fig1}
(d) shows the $I$-$V$ at $T\simeq 0.3$ K, measured at in-plane
magnetic field of $B_{\parallel}=15$ mT. It is seen that $I_c$ is
completely suppressed by a small parallel field, much smaller than
the upper critical fields of BNFA
\cite{Kalenyuk_2017,Prozorov_2014} and Nb \cite{Zeinali_2016}.
Therefore, in such a field we can carefully measure temperature
dependence $R_n(T)$, as shown in Fig. \ref{fig:fig1} (f). The
modest upturn of $R_n$ with decreasing $T$ is quite common for
$c$-axis characteristics of high-$T_c$ superconductors, commonly
associated with a pseudogap \cite{Katterwe_2008}.

\begin{figure*}[t]
    \centering
    \includegraphics[width=0.99\textwidth]{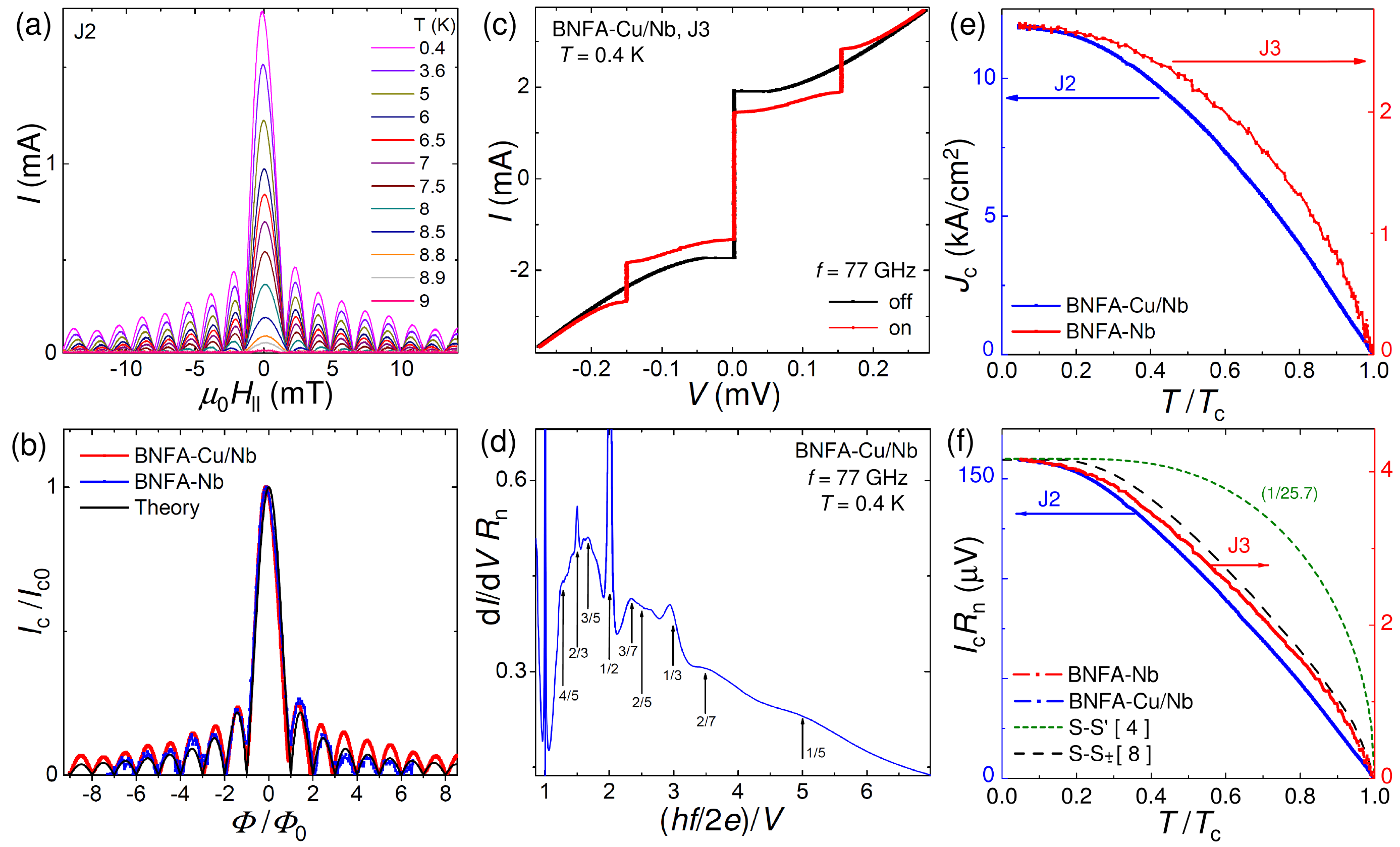}
    \caption{(Color online). (a) Modulation of the critical
    current versus in-plane magnetic field for a BNFA-Cu/Nb
    junctions at different temperatures. (b) Comparison of $I_c$
    versus flux modulation patterns of BNFA-Cu/Nb (red), BNFA-Nb
    (blue) and Fraunhofer pattern (black). (c) $I$-$V$
    characteristics of BNFA-Cu/Nb junction with (red) and without
    (black) microwave radiation. The primary Shapiro step is
    clearly seen. (d) Differential conductance versus inverse
    voltage, demonstrating presence of sub-harmonic Shapiro steps.
    (e) Temperature dependencies of critical current densities of BNFA-Cu/Nb (blue, left
    scale) and BNFA-Nb (red, right scale) junctions. (f) Temperature dependencies of $I_c
    R_n$ products for the same junctions. Note a big difference of $I_c R_n$ values. The dotted black line
    represents a normalized theoretical curve for conventional $s$-wave
    superconductors \cite{AB_1963}. The dashed black line shows a
    simulated dependence for an $s$-$s_{\pm}$ junctions from Ref.
    \cite{Burmistrova_2015}.
}
    \label{fig:fig2}
\end{figure*}

Suppression of $I_c$ by small parallel field is caused by flux
quantization in the junction. Figure \ref{fig:fig2} (a) shows
$I_c(H_{\parallel})$ modulations at different temperatures for a
BNFA-Cu/Nb JJ. Fig. \ref{fig:fig2} (b) represents normalized
$I_c/I_c(0)$ versus flux curves for BNFA-Nb (blue) and BNFA-Cu/Nb
(red) JJ's at low $T$. Both types of JJ's exhibit clear Fraunhofer
modulation depicted by the black line. This is a figure of merit
indicating good uniformity of JJ's \cite{Barone,Likharev}.

Figure \ref{fig:fig2} (c) shows $I$-$V$ curves of a BNFA-Cu/Nb JJ
without (black) and with (red) applied high-frequency
electromagnetic radiation at $f\simeq 74$ GHz at $H=0$ and
$T\simeq 0.4$ K. A clear Shapiro step is seen at $V_1=hf/2e$. Fig.
\ref{fig:fig2} (d) shows the normalized differential conductance
for this $I$-$V$ as a function of $V_1/V$. It reveals numerous
subharmonic Shapiro steps. This indicates the strongly
non-sinusoidal current-phase relation in the JJ
\cite{Ilichev_2004}, which is indeed anticipated for $s$-$s_{\pm}$
JJ's \cite{Burmistrova_2015,Linder_2009}. On the other hand, the
non-sinusoidality may also be caused by the proximity effect in
the Cu/Nb bilayer \cite{Krasnov_1994}.

Thus, our JJ's exhibit clean and clear dc- and ac-Josephson
effects. The high quality of the JJ's together with a good
reproducibility of junction parameters (see the Supplementary
\cite{Suppl}) allows us to investigate genuine characteristics of
composing them superconductors (as opposed to interface defects).
Figs. \ref{fig:fig2} (e) and (f) show temperature dependencies of
(e) the critical current density $J_c$ and (f) the $I_c R_n$
product for both types of junctions. Despite similarities in
behavior, the same BNFA crystal \cite{Suppl}
and fabrication procedure, the two types of JJ's exhibit largely
(almost by two orders of magnitude) different $I_c R_n$ values.
BNFA-Nb JJ's have a very small $I_c R_n \simeq 3~\mu$V
\cite{Kalenyuk_2018}, much smaller than $\Delta/e>1$ mV in both
suuperconductors, while for BNFA-Cu/Nb JJ's $I_c R_n\simeq
200~\mu$V. The difference can be clearly seen in the $I$-$V$
curves from Figs. \ref{fig:fig1} (d) and (e). The reported
remarkable influence of the thin Cu interlayer is the key
observation of this work.

The increase of $I_c R_n$ in BNFA-Cu/Nb JJ's is associated with
the increase of $R_n$. The latter indicates that the interface
transparency, $\beta$, between BNFA and Cu is reduced compared to
BNFA-Nb. Yet, as mentioned above, this does not explain the
increase of $I_c R_n$ because usually $I_c\propto 1/R_n$ and $I_c
R_n$ is independent of $\beta$.

\begin{figure*}[t]
    \centering
    \includegraphics[width=0.9\textwidth]{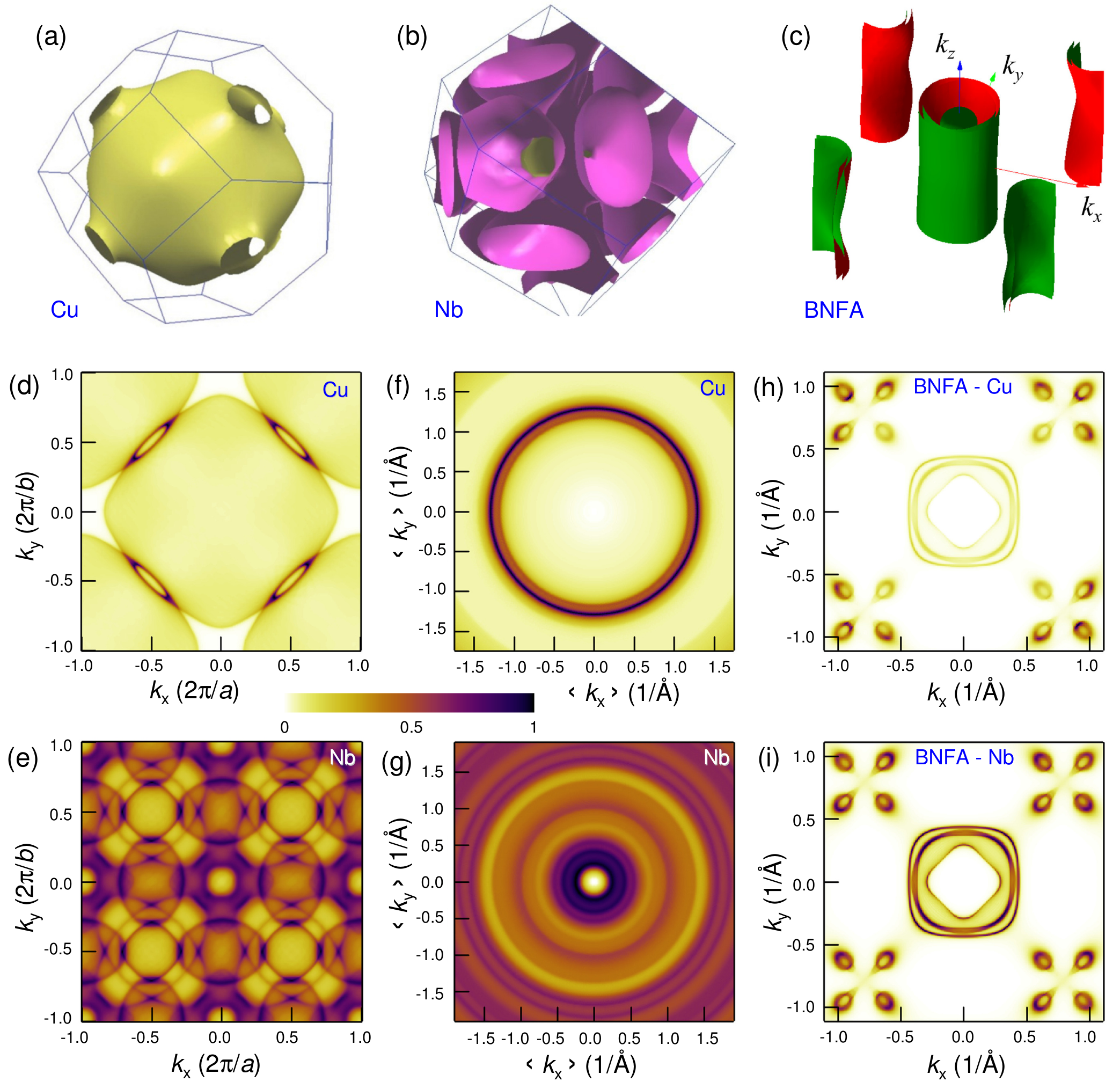}
    \caption{(Color online). (a-c) The Fermi surface topologies in the 1st Brillouin zone for (a) Cu, (b) Nb and (c) BNFA. (d) and
    (e) The $k_z$-integrated projections of Fermi surfaces on the (001) plane for Cu (d) and Nb
    (e). (f) and (g) The same projections averages over the in-plane
    momentum angle, representing the effective ($k_z$-integrated) density-of-states distribution for poly-crystalline Cu (f) and Nb (g) films. (h) and (i)
    Product of $k_z$-integrated Fermi surfaceprojection of BNFAwith effective density-of-states for Cu (h) and Nb (i).
    It can be seen that for BNFA-Cu junctions tunneling
    current flows predominantly into the propeller-like corner bands of BNFA (h). On the other hand, for BNFA-Nb junctions
    the tunneling current is distributed relatively uniformly between central and corner
    bands. }
     \label{fig:fig3}
\end{figure*}

The proximity induced superconducting order parameter in Cu at the
junction interface is $\Phi_N \simeq \beta \Psi_S
\exp(d_N/\xi_N)$, 
where $\Psi_S$ is the order parameter in Nb, $d_N=15$ nm is the Cu
layer thickness and $\xi_N$ is the coherence length in Cu.
According to Ref. \cite{Krasnov_1992}, for thin sputtered Cu films
$\xi_N\simeq 18\sqrt{T_c(Nb)/T}$ (nm), where $T_c(Nb)\simeq 9$ K.
It gives $\xi_N (0.3 K) \simeq 100$ nm, $\xi_N (3 K) \simeq 30$ nm
and $\xi_N (T_c(Nb)) \simeq 18$ nm. Thus, our Cu interlayer is
always thinner than $\xi_N$. Although Cu and Nb films are
deposited without breaking vacuum, the Cu/Nb interface
transparency is modest, $\beta \simeq 0.4$ \cite{Krasnov_1992},
predominantly due to the FS mismatch between Nb and Cu. Thus the
proximity induced order parameter in Cu is smaller than in Nb. For
SINS JJ's (I - insulator, N-normal metal) made of $s$-wave
superconductors, the proximity effect leads to the reduction of
$I_c R_n$ \cite{Golubov_1995}. This is opposite to our
observation. This discrepancy points out that the unconventional
(non-$s$-wave) symmetry of the order parameter in BNFA plays an
essential role. In particular, the extremely small $I_c R_n$ of
BNFA-Nb junctions provides evidence for the sign-reversal
$s_{\pm}$ symmetry in BNFA, due to which supercurrents from bands
with opposite signs of $\Delta$ cancel each other
\cite{Kalenyuk_2018}.

For a more quantitative understanding we consider FS geometries of
involved metals. Figures \ref{fig:fig3} (a-c) show DFT calculated
three-dimensional images of Fermi surfaces for Cu (a), Nb (b)
\cite{FSTableSite,FSTableRef} and BNFA (c) \cite{YareskoPrivat}.
FS of Cu is simple quasi-spherical. The transition metal Nb has a
very complex FS with many small pockets spread over the BZ. BNFA
has two bunches of the FS sheets, the three large quasi-cylinders
in the center and the propeller-like FS's at the corners of the BZ
\cite{ZabolotnyyN2009,Kordyuk2013}. Those bunches are believed to
have opposite signs of $\Delta$ \cite{HirschfeldRPP2011,Wang2011}.

Electron tunneling between two metals usually conserves the
in-plain momentum $\emph{\textbf{k}}_\parallel=(k_x,k_y)$.
Therefore, the total single- electron current can be written as
\begin{equation*}\label{JJjump}
J \propto \oint{T_{12}^2 A_1(k_{z1},\emph{\textbf{k}}_{\parallel})
A_2(k_{z2},\emph{\textbf{k}}_{\parallel})(f_1-f_2)
d\emph{\textbf{k}}_{\parallel} dk_{z1} dk_{z2}},
\end{equation*}
where $T_{12}$ is the tunneling matrix element between initial and
final states, $(k_{z1},k_x,k_y)$ and $(k_{z2},k_x,k_y)$, in the
two electrodes, $A_{1,2}$ are the spectral functions
(momentum-dependent density of states) and $f_{1,2}$ are the
corresponding distribution functions.
The key band-structure-dependent factor is the density of states
projection on the junction plane, which can be integrated
independently
\begin{equation}\label{Nkparallel}
N_i(\emph{\textbf{k}}_\parallel)=\int{A_{i}(k_{zi},\emph{\textbf{k}}_\parallel)
dk_{zi}}, ~(i=1,2).
\end{equation}
Figs. \ref{fig:fig3} (d) and (e) show such projections for Cu and
Nb. The corresponding projection for BNFA is pretty similar to the
pattern, shown in Fig. \ref{fig:fig3} (i).

For comparison with experiment we must take into account the
polycrystalline structure of Cu and Nb electrodes and make an
average with respect to random crystalline orientation. This is
similar to averaging with respect to rotation of $k_{x,y}$ axes.
Figs \ref{fig:fig3} (f) and (g) show thus averaged projections,
$<N_i(k_x,k_y)>$, for polycrystalline Cu and Nb. The key
difference is that due to the quasi-spherical FS of Cu, the
polycrystalline density of states projection keeps the circular
shape with the radius given by the Fermi momentum. To the
contrary, averaging for multi-band polycrystalline Nb leads to a
more uniform distribution of the density of states.

Figs. \ref{fig:fig3} (h) and (i) show a product of the density of
state projections (h) $<N_{Cu}(k_x,k_y)>N_{BNFA}(k_x,k_y)$ for
BNFA-Cu/Nb and (i) $<N_{Nb}(k_x,k_y)>N_{BNFA}(k_x,k_y)$ for
BNFA-Nb junctions. It gives a hint about contribution of the two
BNFA bands in electrical current through the junction. For BNFA-Nb
JJ's both BNFA bands participate approximately equally due to
fairly uniform distribution of $<N_{Nb}(k_x,k_y)>$ in the BZ
projection, Fig. \ref{fig:fig3} (g). To the contrary, the highly
non-uniform, circular-shape $<N_{Cu}(k_x,k_y)>$, Fig.
\ref{fig:fig3} (f), blocks tunneling into the central band of
BNFA.

For calculation of supercurrent, $A_i$ should be replaced by
$A_i\Psi_i$, where $\Psi_i$ is the superconducting order parameter
in the corresponding metal. For Nb and Cu/Nb with $s$-wave order
parameter $\Psi$ is just a number. However, for the unconventional
two-band superconductor BNFA, which likely has the $s_{\pm}$
symmetry, $\Psi$ changes sign between central and corner bands.
For BNFA-Nb JJ's with similar transport contribution of the two
bands this leads to an almost complete cancellation of the total
supercurrent \cite{Kalenyuk_2018}. However, for BNFA-Cu/Nb JJ's
the cancellation is much smaller because tunneling from central
bands is suppressed.
Therefore, such analysis qualitatively explains larger values of
both $R_n$ and $I_c R_n$ in BNFA-Cu/Nb junctions.


To conclude, we fabricated and studied high-quality Josephson
junctions between an iron-pnictide superconductor
Ba$_{1-x}$Na$_x$Fe$_2$As$_2$ and either a conventional low-$T_c$
superconductors Nb or a Cu/Nb bilayer. Remarkably, we observed
that addition of a very thin (15 nm) Cu interlayer changes
drastically junction properties and, in particular, increases the
$I_c R_n$ product by almost two orders of magnitude. The latter is
opposite to expectations for proximity-coupled junctions made of
conventional $s$-wave superconductors \cite{Golubov_1995}. This
counterintuitive result adds to evidence for the unconventional
$s_{\pm}$ symmetry of the order parameter in the pnictide. The
phenomenon is explained qualitatively taking into account
particular Fermi surface geometries of involved metals. It is
shown that the multi-band structure of Nb leads to similar
contributions of both pnictide bands into electron transport,
which due to the sign-reversal $s_{\pm}$ superconducting order
parameter in the two electronic bands of the pnictide, leads to
the cancellation of the total supercurrent and results in a very
small $I_c R_n \simeq 3~\mu$V \cite{Kalenyuk_2018}. To the
contrary, the simple quasi-spherical Fermi surface of Cu supports
tunneling predominantly from only one band, avoiding the
supercurrent cancellation and resulting in much larger $I_c
R_n\simeq 200~\mu$V. Our results indicate that unlike for
junctions made of conventional $s$-wave superconductors, for
junctions with unconventional sign-reversal superconductors the
Fermi surface geometry plays a crucial role. This provides a new
tool for phase sensitive studies of such materials and could
probably explain some of reported variations of $I_c R_n$ values
in pnictide JJ's
\cite{Kalenyuk_2018,Burmistrova_2015,Seidel_2017}. The reported
material-dependence of tunneling into pnictide superconductors can
be used for optimization and conscious engineering of
pnictide-based Josephson junctions.

\begin{acknowledgements}

The work was supported by the National Research Foundation of
Ukraine (project 2020.02/0408) and the Russian Science Foundation
(Grant No. 19-19-00594). The paper was accomplished during a
sabbatical period of V.M.K. at MIPT. We are grateful to
A.N.~Yaresko for providing the results of DTF calculations,
V.V.~Zabolotnyy for help with FS visualization, S.~Aswartham,
S.~Wurmehl and B.~B\"{u}chner for providing BNFA crystals.

\end{acknowledgements}


\bibliographystyle{PRL}

\end{document}